\documentstyle[prl,aps,preprint]{revtex}
\begin{document}

\title {Exact dynamical response of an N-electron quantum dot\\ 
subject to a time-dependent potential} % 
\author{Simon C. Benjamin and  Neil F.
Johnson} 
%\begin{instit} 
\address{Physics Department, Clarendon
Laboratory, Oxford University, Oxford OX1 3PU, England} 
%\end{instit}

\maketitle

\begin{abstract}

We calculate analytically the exact dynamical response of a droplet of
$N$ interacting electrons in a quantum dot with an arbitrarily
time-dependent parabolic confinement potential $\omega(t)$ and a
perpendicular magnetic field. We find that, for certain frequency ranges, a
sinusoidal perturbation acts like an {\em
attractive} effective interaction between electrons. In the 
absence of a time-averaged confinement potential, the $N$ electrons can bind
together to form a stable,
free-standing droplet. 

\end{abstract}  \vskip 0.2in  


\vskip 0.2in

\newpage
%\narrowtext

Quantum dots have attracted much interest recently, both from a pure and
applied viewpoint.
According to the dot fabrication, the confinement length
scales in the three spatial directions can be quite
different yielding quasi one-, two- or three-dimensional dots 
\cite{neil}. 
In
addition the number of electrons $N$ in the dot can be reduced down to
the few-electron limit. Most theoretical quantum dot research has been
concerned with the linear response of the resulting $N$-electron,
low-dimensional system. External, time-dependent electric fields are
typically treated as small perturbations which merely give rise to 
transitions between the eigenstates of the unperturbed dot. 
Given the technological possibilities for preparing stronger fields, it
is interesting to consider the effects of larger, time-dependent
perturbations on such dots. For example, a confinement potential with
sinusoidal time-dependence could be created by applying an a.c. bias to the
electrodes defining the dot in a heterostructure sample.  Unfortunately,
the  quantum-mechanical problem of $N$ interacting electrons in a dot
subject to an arbitrarily strong, time-varying perturbation is too
complicated to solve in general, even numerically.

Here we provide an analytically solvable model for the dynamical
response of an $N$ electron dot with an arbitrarily strong,
time-varying confinement  potential, in the presence of an arbitrarily    
strong magnetic field $B$.  
The analytic tractability of the model is made
possible through a combination of a parabolic form for the
dot confinement potential and an inverse-square
electron-electron interaction potential ($1/r^n$ with $n=2$). The
parabolic confinement is a reasonable
approximation for many semiconductor quantum dot samples \cite{neil}.
The
true repulsive interaction between  electrons in the dot is
likely to be better fit by $n\sim 1$ at small $r$ and $n\sim 3$ at
large $r$ due to image charge effects in neighboring electrodes
\cite{Maksym}; however the general features of our results with $n=2$
for all $r$
should be qualitatively correct. Here we focus on the usual
quasi-two-dimensional dot; however, most of our formal results
can be generalized to both quasi-one- and quasi-three-dimensional
dots at $B=0$. 

 The time-dependent
Schrodinger equation for the dot with a
time-dependent confinement in the xy-plane, subject to a
constant $B$ field applied along z, is given within
the effective-mass approximation by $H(t)\Psi(t)=i\hbar
\frac{\partial}{\partial t}\Psi(t)$ with \begin{equation} H(t) =\sum_i
({{\bf p}_i^2\over {2m^*}}  + {1\over 2}m^*\omega^2(t)|{\bf r}_i|^2 +
{\omega_c\over 2}l_i) + \sum_{i<j} \frac{\xi}{|{\bf r}_i-{\bf r}_j|^2} \
. \end{equation} The momentum, position and z-component of angular
momentum of the $i$'th electron are given by ${\bf p}_i$,
${\bf r}_i$  and $l_i$. The cyclotron frequency is $\omega_c$ (N.B. we
can only solve with $\omega_c\neq 0$ for a two-dimensional dot). The
spin part of the Hamiltonian is time-independent and therefore decouples
for all $t$. In order to include both transient and steady-state
responses, we consider the time-dependent dot potential to be turned on
at time $t=0$: 
\begin{equation}
\omega^2(t)=\cases{\omega_0^2(B)=\omega_0^2+\omega_c^2/4& for $t\leq 0$\cr
f(t)& for
$t>0\ \ $\cr} 
\end{equation} where $\omega_0$ is the
characteristic frequency of the parabolic confinement potential for
$t\leq 0$, and $f(t)$ has arbitrary functional
form and magnitude.

For $t<0$, $\omega^2(t)$ is time-independent. The problem is treated in
Ref. \cite{PRL}; here we review the essential results. Standard
Jacobi coordinates are employed, ${\bf X}_i$ ($i=0,1,\dots ,N-1$) where  
${\bf X}_0=\frac{1}{N}\sum_j {\bf r}_j$ (center-of-mass),  ${\bf
X}_1=\sqrt{\frac{1}{2}}({\bf r}_2-{\bf r}_1)$,   ${\bf
X}_2=\sqrt{\frac{2}{3}}(\frac{({\bf r}_1+ {\bf r}_2)}{2}-{\bf r}_3)$ 
etc. together  with their conjugate momenta ${\bf P}_i$.  The
center-of-mass motion decouples, $H=H_{\rm CM}({\bf X}_0)+ H_{\rm
rel}(\{{\bf X}_{i>0}\})$, hence $E=E_{\rm CM}+E_{\rm rel}$ and
$\Psi=\psi_{\rm CM}\psi_{\rm rel}$. The exact eigenstates $\psi_{\rm CM}$
of $H_{\rm CM}$ and eigenenergies $E_{\rm CM}$ are identical to those of
a single particle in a parabolic potential. The
non-trivial problem  is to solve the relative motion equation  $H_{\rm
rel}\psi_{\rm rel}=E_{\rm rel}\psi_{\rm rel}$. We transform the relative
coordinates  $\{{\bf X}_{i>0}\}$ to standard hyperspherical coordinates:
${\bf X}_i=r(\prod_{j=i}^{N-2}{\rm cos}\alpha_{j+1}){\rm
sin} \alpha_{i}  e^{i\theta_i}$ with $r\geq 0$ and
$0\leq\alpha_i\leq\frac{\pi}{2}$ ($\alpha_{1}=\frac{\pi}{2}$).
Physically, the hyperradius $r$ is just the root-mean-square
electron-electron separation. The exact eigenstates of $H_{\rm rel}$
have the form $\psi_{\rm rel}=R(r)F(\tilde\Omega)$ where $\tilde\Omega$
denotes the $(2N-3)$ hyperangular  $\{\theta,\alpha\}$ variables; $R(r)$
and $F(\tilde\Omega)$ are  solutions of the hyperradial and
hyperangular  equations respectively. 

For $t>0$, the separation of the relative motion is still exact, i.e.
$\Psi=\psi_{\rm CM}(t)\psi_{\rm rel}(t)$. The hyperangular equation
is independent of $\omega(t)$, hence $F(\tilde\Omega)$ remains
time-independent. The time-dependence of $\Psi$ is
only contained in the center-of-mass and hyperradial parts. Consider
explicitly $\psi_{\rm rel}(t)$ and hence the hyperradial part $R(r,t)$; the
solution for $\psi_{\rm CM}(t)$ is exactly
analogous. Following Ref. \cite{Camiz} we construct the generating function
\begin{equation} g(z,r,t)\equiv\sum_{n=0}^\infty R_{n}(r,t) z^n \end{equation}
where
$R_{n}(r,t)$ are the solutions of the time-dependent hyperradial
equation.  Because the dot potential is constant for $t\leq 0$, we can
employ the static $R_n(r)$ from Ref.\cite{PRL} to obtain \begin{equation}
g(z,r,t\leq 0)=\sum_{n=0}^\infty \big(\frac{r}{l_0}\big)^\gamma L_n^{\gamma+N-2}
\big(\frac{r^2}{l_0^2}\big) e^{-\frac{r^2}{2l_0^2}} z^n 
\end{equation}
where $L$ denotes the Laguerre polynomial and
$l_0^2=\hbar(m^*\omega_0(B))^{-1}$. 
The parameter $\gamma$ is determined by the $\omega_0$, $B$ and $t$ -independent
 hyperangular equation, which does not admit complete 
exact solution. Fortunately, we do not need any further knowledge of the
properties of $\gamma$ for the analysis in this paper.
Equation (4) can be written in closed form using a known identity
\cite{tables}
\begin{equation} g(z,r,t\leq 0)= \big(\frac{r}{l_0}\big)^{\gamma}
 e^{{{z+1}\over{2(z-1)}}(\frac{r}{l_0})^2}(1-z)^{-(a+1)} \end{equation}
where $a=\gamma+N-2$. We make the ansatz \begin{equation} g(z,r,t>0)=
\alpha(z,t) ({r\over l_0})^{\gamma} e^{\alpha'(z,t) r^2} \end{equation}
which can be shown to satisfy the time-dependent hyperradial equation
\begin{equation} \big(\frac{\partial^2}{\partial r^2}+ 
\frac{2N-3}{r}\frac{\partial}{\partial
r}-\frac{\gamma(\gamma+2N-4)}{r^2}-
({{m^*}\over\hbar})^2\omega^2(t)r^2)g(z,r,t)=-{{2im^*}\over\hbar}
\frac{\partial}{\partial t}g(z,r,t) \end{equation}  and the $t=0$
boundary condition in Eq. (5), provided \begin{equation}
\alpha(z,t)=[\eta(t)]^{-(a+1)}exp[2i\theta(t)(a+1)](1-z\
exp[2i\theta(t)])^{-(a+1)} \end{equation} and \begin{equation}
\alpha'(z,t)={{im^*}\over{2\hbar}}({{\eta\dot(t)}\over{\eta(t)}}
-2i\theta\dot(t)(1-z\ exp[2i\theta(t)])^{-1}) \end{equation} where
$\eta(t)=|\eta(t)|e^{i\theta(t)}$ solves the {\em classical}
one-dimensional oscillator 
\begin {equation} \eta\ddot(t)+f(t)\eta(t)=0
\end {equation} with boundary conditions $\eta(0)=1$ and
$\eta\dot(0)=-i\omega_0(B)$. We may then expand $g(z,r,t)$ using the
relevant
identity \cite{tables}, and compare coefficients of $z^n$ with the defining
Eq.
(3) to obtain the desired time-dependent  wavefunctions (unnormalized):
\begin{equation} R_n(r, t>0)= |\eta(t)|^{1-N} y^\gamma exp[i(
\theta(t)(2n+a+1)+{{y^2}\over{4\omega_0(B)}}{d\over{dt}}|\eta(t)|^2 )] 
e^{-{1\over2}y^2}L_n^a(y^2) \end{equation} where
$y\equiv{r\over{|\eta|\l_0}}$. Solving Eq. (10) hence provides a complete
description for the evolution of the initially stationary hyperradial
state $R_n$ for $t>0$. Equation (11) together with the exactly
analogous expression for the center-of-mass entirely determines the
time-evolution of the total wavefunction $\Psi$.

First we consider a step-like perturbation \cite{magnet} (see Fig. 1(a))  
\begin{equation}
\omega^2(t)=\cases{\omega_0^2& for $t\leq 0$\cr
f(t)=\omega^2_{\infty}-{{w^2_{\infty}-\omega_0^2}\over{(\Omega t+1)^2}}&
for $t>0\ \ $\cr} \ \ .\end{equation} Increasing $\Omega$ reduces the
time period over which the change occurs. The
solution to Eq. (10) for $t>0$ is \begin{equation}
\eta(t)=(\Omega t+1)^{1\over2}\{AJ_v[\omega_\infty(t+\Omega^{-1})]+
BN_v[\omega_\infty(t+\Omega^{-1})]\} \end{equation} where $J_v$ and
$N_v$ are Bessel functions of the first and second kind respectively,
$v=({{(\omega_\infty^2-\omega_0^2)}\over{\Omega^2}}+ {1\over
4})^{1\over2}$, and $A$ and $B$ are complex constants chosen to satisfy
$\eta(0)=1$ and
$\eta\dot(0)=-i\omega_0$. This complex function is displayed
in Fig. 1(b), the inset on the right showing the view when projected on
the complex $\eta$ plane. For $t<0$ the value of $\eta(t)$ lies on
the unit circle in the complex plane, and describes a simple helix around
the time axis as $t$ varies. For $t>0$ the function describes an
ellipsoidal path in the complex $\eta$ plane, the ellipsoid lying
entirely within the unit circle; the eccentricity of the ellipsoid is
determined by the sharpness of the transition from $\omega_0$ to
$\omega_\infty$, i.e. by $\Omega$. Since $R_n(r,t)$ (Eq. 11) depends on
$y\equiv{r\over{|\eta|\l_0}}$, the ellipsoidal path of $\eta(t)$ implies
that the total wavefunction will oscillate for all $t>0$. Figure 1(c)
shows the time-dependence of the expectation value  of the
hyperradius $\bar r$ (r.m.s. electron-electron separation). Remarkably,
this quantity oscillates for all $t>0$ regardless of the values of
$\omega_0$, $\omega_\infty$ and $\Omega$. Changing $\Omega$ alters the
oscillation amplitude but not the period. In order to obtain  a large
amplitude of oscillation, $\Omega^{-1}$ must be of comparable order to
$\omega_0^{-1}$; if  $\hbar\omega_0\sim 1$ meV,
$\Omega^{-1}$ should be of the order of picoseconds.

Second we consider a sinusoidal perturbation \cite{magnet} (see Fig. 2(a)):
\begin{equation} \omega^2(t)=\cases{\omega_0^2& for $t\leq 0$\cr
f(t)=\omega^2_0-\omega_1^2(1-{\rm cos}(2\Omega t))& for $t>0\ \ $\cr}
\end{equation} The solution to Eq. (10) for
$t>0$ is a Mathieu function: 
\begin{equation} \eta(t)=A
e^{\mu\Omega t}\sum_{n=-\infty}^{\infty} c_{2n} e^{2in\Omega t}+
        B e^{-\mu\Omega t}\sum_{n=-\infty}^{\infty} c_{2n} e^{-2in\Omega
t} \end{equation} where $\mu$ and $\{c_{2n}\}$ are determined by 
simultaneous equations (see in Ref. \cite{tables}), and $A$ and $B$ are
complex
constants chosen to satisfy $\eta(0)=1$ and $\eta\dot(0)=-i\omega_0$.
Figure 3
shows $\mu$ as a function of $\omega_0$,
$\omega_1$ and $\Omega$. The circle in Fig. 3 indicates the
parameters
for Fig. 2(a). In certain regions
(white) of
the parameter space
$\mu$ is purely imaginary while in others (dark) it has  a real
part. This real part indicates that the classical particle is resonating
with the oscillating dot; the particle's  oscillations become
infinitely large as $t\rightarrow\infty$. The corresponding effect  on
the quantum mechanical system, which depends on time only through
$\eta(t)$, will be an increase in energy as $t\rightarrow\infty$ and a
decrease in localization of $R(r,t)$ and hence $\psi_{\rm rel}$. This
spreading in $\psi_{\rm rel}$ implies an increase in the average
electron-electron separation, and will lead to electrons escaping from
any realistic dot having a finite depth.
 For small $\omega_1$,  $\mu$ has a real component only when the
perturbing frequency $\Omega$ is equal to  an integer {\em fraction}
${1\over n}$ of the dot potential $\omega_0$. Note our
treatment is exact for any amplitude $\omega_1^2$. The theory
remains valid even when $\omega_1^2>{1\over2}\omega_0^2$, in which case
the dot becomes repulsive (i.e. an `anti-dot') for part of each
oscillation as shown in Fig 2(a).
The region of Fig. 3 below the line $\bar{\omega}=\omega_1$ corresponds to
such a system. Stable zones persist here, thus {\em  a dot may periodically
become repulsive and yet confine its electrons for all time}. Indeed in the limit
$\omega_1=\omega_0$ (the bottom edge in Fig. 3) {\em the
$N$-electron droplet  may
remain localized for all time even though the time-averaged confinement
potential is zero} (i.e. $\bar{\omega}^2=\omega_0^2-\omega_1^2=0$). 
The sinusoidal perturbation in these white, stable zones
generates an effective {\em attractive} 
interaction
between electrons which competes with the electrostatic repulsion. At
${\bar\omega}=0$ the result is
a stable, free-standing $N$-electron plasma droplet. From the inset
in Fig. 3 we see that, for $\bar \omega=0$, the range of
the parameter ${\omega_1\over
\Omega}$  over which the system is stable (i.e. white zones) is
small but finite for larger ${\omega_1\over
\Omega}$; we note that there an {\em infinite} number of such stable zones 
which 
are distributed along the entire $\bar\omega=0$ axis.

Varying only $\Omega$ 
corresponds
to moving along a radial line in Fig. 3. Consider the
indicated line
$\bar{\omega}=\omega_1$; moving out along this line
switches the system between unstable states (dark zones -- system absorbs
energy from the perturbing field) and stable states (white zones -- no
net absorption). This pattern seems to persist arbitrarily far along
this radial line, hence an experiment to study this stability effect
could be performed at {\em any} frequency   $\Omega$ which is
convenient, irrespective of $\omega_0$. This observation 
is encouraging experimentally 
since the frequency $\omega_0$ is typically
high ($~10^{12}$ Hz for 
$\hbar\omega_0=1$meV).

We now focus on `stable'
solutions (white zones in Fig. 3). Figure 2(b) shows
the time-variation of  the r.m.s.
electron-electron separation ${\bar r}(t)$ for a sinusoidal
perturbation (Fig. 2(a)); it
seems to show a chaotic irregularity (N.B. the amplitude is always finite
since we are in the stable regime). The function ${\bar r}(t)$ is
actually the square-root of a sum of cosine functions; Fig. 2(c) shows the
amplitudes and frequencies of the  dominant terms in this sum. They are
made up of two distinct groups: one at frequencies $2n\Omega$ and another
at $(2n+1)\Omega\pm \beta\Omega$. Two lines from the second group are
labeled as an example. The value of $\beta$ may be found from Fig. 3; for
the present choice of parameters (marked
by a circle in Fig. 3) 
$\beta=2.35$. Figure 3 shows that when the perturbation amplitude
$\omega_1$ is
small then 
$\beta\approx{\bar{\omega}\over \Omega}$. Thus
for weak-to-moderate perturbations, the second group of peaks in
the frequency spectrum lie at sums and differences  of $\bar{\omega}$ and
the
driving frequency
$\Omega$, i.e. at $(2n+1)\Omega\pm \bar\omega$.
The system can hence exhibit significant frequency mixing and harmonic
generation. Figure 3 shows how $\beta$ (and hence the
spectrum) changes as we move to the strongly non-linear regime 
($\omega_1\approx\bar{\omega}$). The parameters for Fig. 2
correspond to a strong perturbation ($\omega_1^2=0.57\omega_0^2$) which 
causes
$\Omega\beta$ to deviate from $\bar{\omega}$ by 
$16\%$. 

In summary, we have presented a method for finding the exact dynamical
response of an
$N$ electron droplet. Our first example shows that a droplet
subjected to a sudden increase in confinement will exhibit novel size
oscillations associated with dynamical breathing modes {\em for all time}. Our
second example shows that a  sinusoidally varying confinement generates strongly
non-linear size oscillations whose stability depends non-trivially on the
confinement parameters. Electrons may remain localized {\em even in the absence
of a time-average confinement potential}, thereby offering a possible new method
of quantum dot fabrication. This plasma droplet could be moved around in
the plane under the influence of a small, static in-plane electric field.
In both examples the size oscillations are strong in that they involve {\em all}
the electrons in the dot, and should therefore be readily observable
experimentally.

We thank Luis Quiroga and Nikos Nicopoulos for useful
discussions. S.C.B. is supported by an EPSRC studentship.

\newpage

\bigskip

\centerline{\bf Figure Captions}

\bigskip

\noindent Figure 1. (a) Step-like perturbation $\omega^2(t)$. (b)
$\eta(t)$; inset shows the  projection onto the complex $\eta$ plane.
(c) The r.m.s. electron-electron separation ${\bar r}$(t).

\bigskip

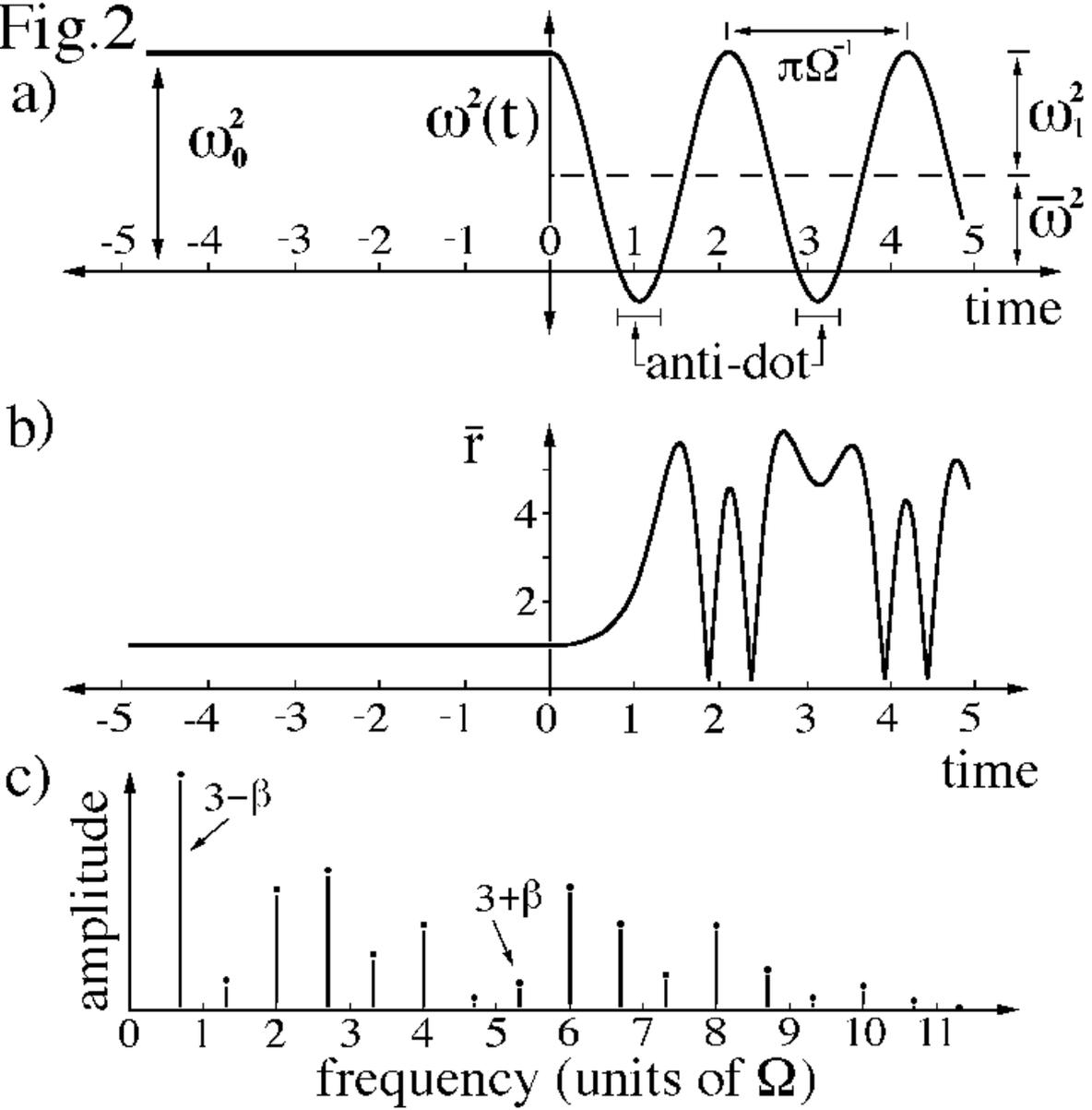
\noindent Figure 2. (a): Sinusoidal perturbation $\omega^2(t)$; 
$\bar\omega^2\equiv\omega_0^2-\omega_1^2$.  
(b) The r.m.s. electron-electron separation ${\bar r}$(t).
(c) Frequency (Fourier) composition of ${\bar r}^2(t)$. 
\bigskip

\noindent Figure 3. Top: Contour plot showing 
$\mu$ as a function of perturbation
parameters;
$\mu=Re\{\mu\}+i\beta$. 
White regions:
$Re\{\mu\}=0$ and contours are lines of constant $\beta$; on
$\omega_1=0$
axis $\beta={\bar\omega\over\Omega}$. Dark regions:
$Re\{\mu\}>0$ and contours are lines of constant $Re\{\mu\}$ in
increments
of 0.2, as shown by key.


\begin{thebibliography}{99}
\bibitem{magnet} The B-field can be trivially introduced by replacing
$\omega_0$ by $\omega_0(B)$.
\end{thebibliography}
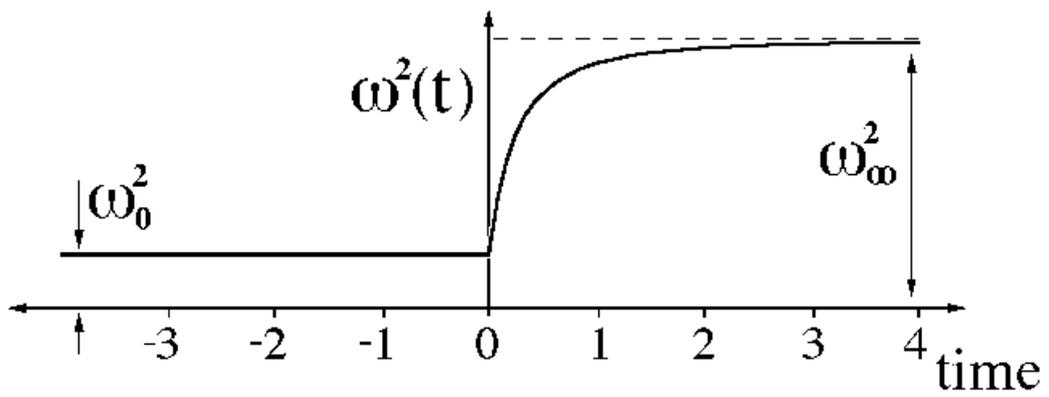
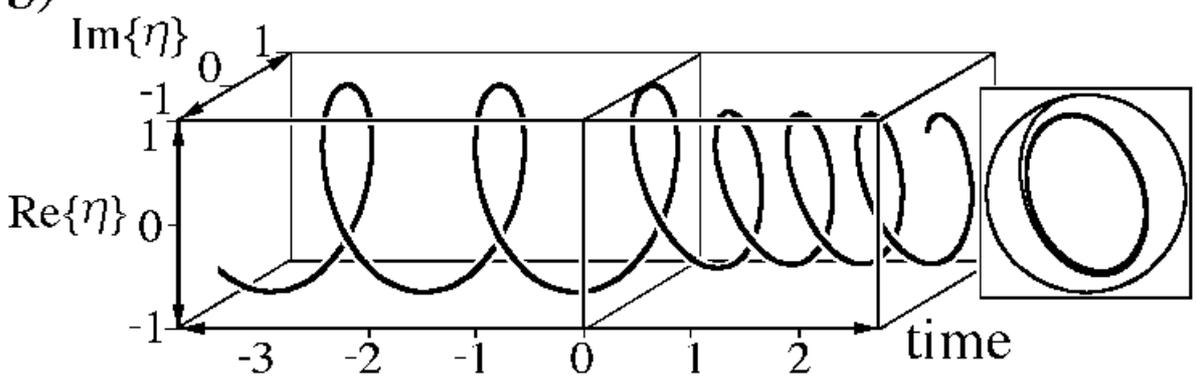
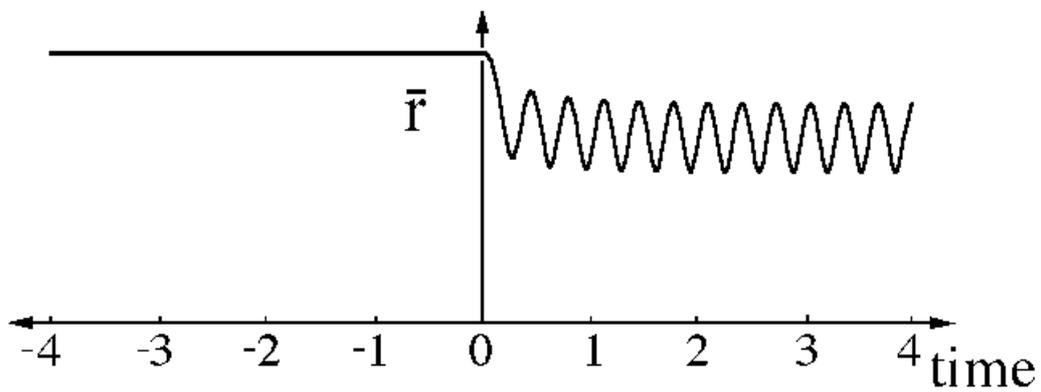
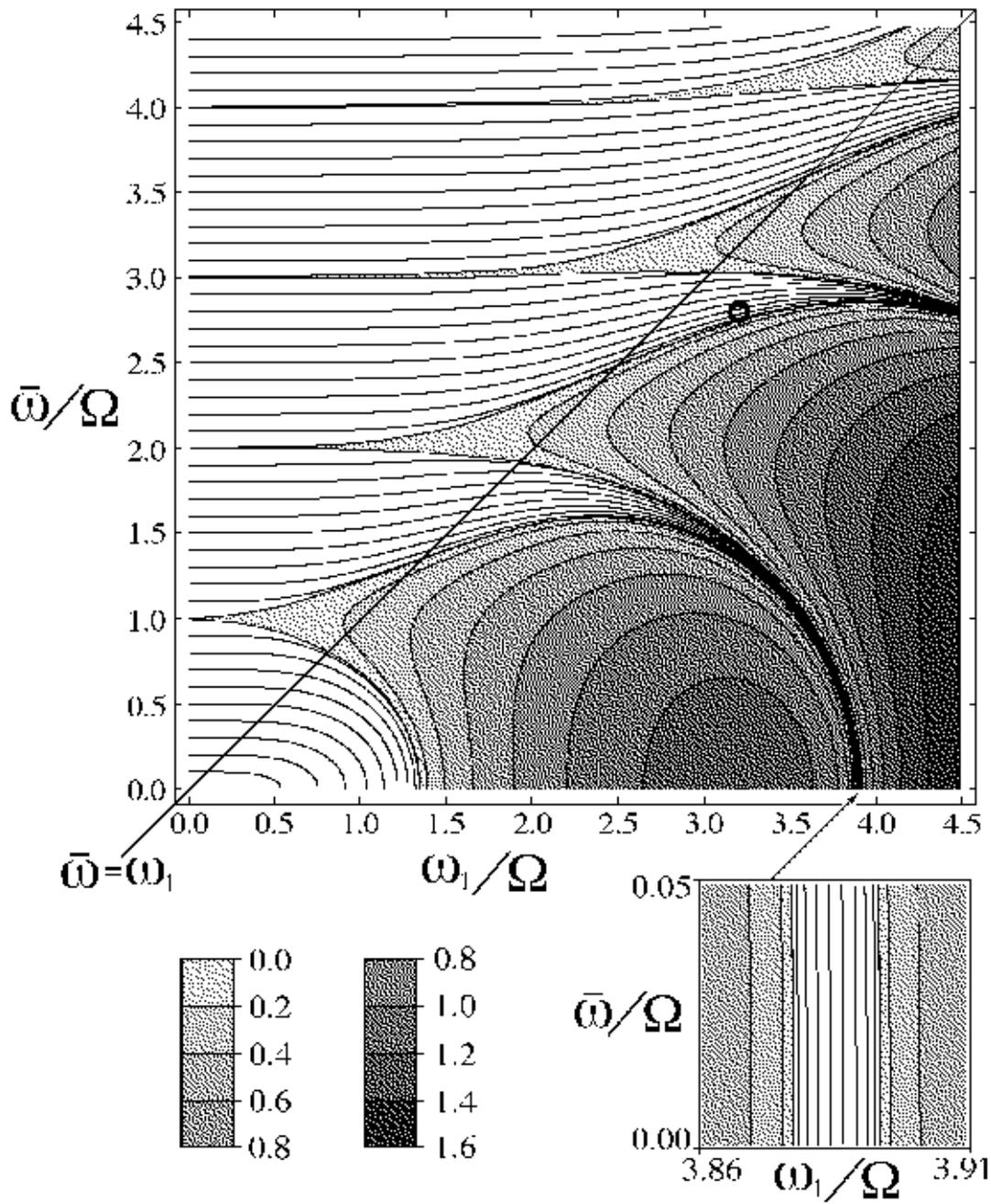
\end{document}